\documentclass{elsart}

\usepackage{amssymb}

\begin{document}

\begin{frontmatter}



\title{Quantum features in statistical observations\\   
    of ``timeless'' classical systems}


\author{H.-T. Elze}
\ead{thomas@if.ufrj.br}
\address{Instituto de F\'{\i}sica, Universidade Federal do Rio de Janeiro \\ 
C.P. 68.528, 21941-972 Rio de Janeiro, RJ, Brazil}

\begin{abstract}
We pursue the view that quantum theory may be an emergent structure related   
to large space-time scales. In particular, we consider classical Hamiltonian 
systems in which the intrinsic proper time evolution parameter is related through 
a probability distribution to the discrete physical time. This is motivated by  
studies of ``timeless'' reparametrization invariant models, where discrete physical 
time has recently been constructed based on coarse-graining local observables. 
Describing such deterministic classical systems with the help of path-integrals, 
primordial states can naturally be introduced which follow unitary quantum 
mechanical evolution in suitable limits.
\end{abstract}

\begin{keyword}
reparametrization invariance \sep discrete time \sep emergent quantum theory
\PACS 03.65.Ta \sep 04.20.-q \sep 05.20.-y
\end{keyword}
\end{frontmatter}
\begin{center} 
{\it Dedicated to Constantino Tsallis on occasion of his 60th birthday.}
\end{center}
\section{Introduction}
Since its very beginnings, there have been speculations on the possibility of 
deriving quantum theory from more fundamental dynamical structures, possibly deterministic 
ones \cite{I,II}. Famous is the discussion by Einstein, Podolsky and Rosen, interpreted as 
the need for a more complete fundamental theory. 
However, just as numerous have been attempts to prove 
no-go theorems prohibiting exactly such ``fundamentalism'', culminating in the studies of Bell. The EPR paradox as well as the Bell inequalities have come under experimental 
scrutiny in recent years, confirming the predictions of quantum mechanics in laboratory experiments on scales very large compared to the Planck scale.   
  
However, to this day, the feasible experiments cannot rule out the possibility that quantum mechanics emerges as an effective theory only on sufficiently large scales and can indeed  
be based on more fundamental models.     

Motivated by the clash between general relativity and quantum theory, 't\,Hooft has 
strongly argued in favour of model building in this context \cite{tHooft01,tHooft02}. 
Various further arguments for deterministically induced quantum features have recently been proposed, 
for example, in Refs.\,\cite{Wetterich02,Vitiello01,Mueller,Smolin,Adler}, in the context of statistical systems, of  
considerations related to quantum gravity, and of matrix models, respectively.   
-- Here I report on a large class of 
classical deterministic systems which show emergent quantum mechanical features \cite{I}.  
This is based on recent work on time-reparametrization invariant models \cite{ES02,E03}, 
where discrete phyical time has been constructed. 
Essential aspects of what has been presented in my talk will be summarized, 
while the details may be found in the references. 

\subsection{Discrete physical time in ``timeless'' classical models}  
Analogous to common gauge theories, e.g. the standard model of particle physics, 
reparametrization invariant systems are invariant under a kind of gauge transformation. 
In the case of diffeomorphism invariant theories, such as general relativity 
or string theory, this amounts to invariance under general coordinate transformations. 
Considering time-reparametrization invariance only, this means  
invariance of the dynamics under arbitrary transformations:
\begin{equation}\label{timerepara} 
t\;\longrightarrow t'\;,\;\;\mbox{with}\;\;t\equiv f(t') 
\;\;, \end{equation}    
with $f$ monotonous and differentiable. 
The corresponding constrained Lagrangian dynamics is discussed in Refs.\,\cite{ES02,E03} 
for the respective models. 
 
Similarly as Gauss' law in electrodynamics, for example, an important consequence of time-reparametrization 
invariance is the (weak) constraint that the Hamiltonian has to vanish (on the solutions of the equations 
of motion).
Since the Hamiltonian commonly is the generator of time evolution, this 
has led to name such systems ``timeless''. A Newtonian external time parameter does not exist. 
Problems arise when trying to quantize such a system, since the standard Schr\"odinger 
equation does not exist. Thus, the  
Wheeler-DeWitt equation, $\widehat H|\psi\rangle =0$, epitomizes the intrinsic problems of quantum gravity.  
  
Numerous approaches have been tried to resolve this (in)famous ``problem of time''. 
In distinction to others, I insist on a {\it local} description. 
Still, for a particle with 
time-reparametrization invariant dynamics, be it relativistic or nonrelativistic, 
one can define quasi-local observables 
which characterize the evolution in a gauge invariant way \cite{ES02,E03}.    
Essentially, some of the degrees of freedom of the system are employed to trigger a localized 
``detector'', which can be defined invariantly. It amounts to attributing to 
an observer the capability to count discrete events. Then,
the detector counts present an observable 
measure of {\it discrete physical time}.  

This result can be understood differently by noting that a Poincar\'e section 
is invoked here, which reflects an ergodic 
if not periodic aspect of the dynamics -- quite analogous to a pendulum 
which triggers a coincidence counter each ``time'' it passes through its equilibrium 
position.   
Reparametrization invariance strongly limits the information which can be extracted from it  
with respect to a complete trajectory. This is the reason that physical time 
based on local observations (clock readings) necessarily is discrete.  
  
The possibility of a fundamentally discrete time (and possibly other discrete 
coordinates) has been explored before, ranging from an early realization of Lorentz 
symmetry in such a case to detailed explorations of its consequences 
and consistency in classical mechanics, quantum field theory, and general relativity. 
So far, however, no classical physical models implying such discreteness were proposed.  
{\it Quantization as an additional step} -- resulting in discreteness of coordinates in 
some cases -- has always been performed as usual. 

\subsection{Does discrete time induce quantum mechanical features?}  
There are indications for a qualified `Yes', answering this question.  
The findings of Refs.\,\cite{ES02,E03} suggested that those discrete-time models 
can be mapped on a cellular automaton studied by 't\,Hooft \cite{tHooft01,tHooft02}. Employing 
the algebra of $SU(2)$ generators, 
it has been shown that these models actually reproduce the quantum mechanical harmonic 
oscillator in a large-scale limit. 
The Ref.\,\cite{I} presents an attempt to show more generally 
that due to inaccessability of globally complete information on trajectories of the system, 
the evolution of remaining degrees of freedom appears as in a quantum mechanical model 
when described in relation to the discrete physical time.     
  
We may call this ``stroboscopic'' quantization: when a continous 
physical time is not available but a discrete one is -- like reading an analog clock under 
a stroboscopic light -- then states of the system which fall in between subsequent clock 
``ticks'' cannot be resolved. (Of course, evolution in the unphysical parameter time 
is continous in the constrained Lagrangian models we refer to.) Such unresolved states  
form equivalence classes which can be identified with primordial 
Hilbert space states \cite{tHooft01,tHooft02,Wetterich02,ES02,E03}. The residual dynamics then  
leads the evolution of these states through discrete steps. Under favourable circumstances, 
this results in unitary quantum mechanical evolution.   

Presently, incorporating the discreteness of time is simplified by 
relating the physical time $t$ via a probability distribution $P$ to the proper time $\tau$ of 
the equations of motion: 
\begin{equation}\label{P} 
P(\tau ;t)\equiv P(\tau -t)\equiv\exp\Big (-S(\tau -t)\Big )\;\;,\;\;\;
\int\mbox{d}\tau\;P(\tau ;t)=1
\;\;. \end{equation}  
Note that the almost perfect clock described in this way does not age with 
physical time.
Explicit examples of this can be found in Refs.\,\cite{ES02,E03}, 
when the clock degrees of freedom evolve independently of the rest of the system, apart 
from the Hamiltonian constraint.    
Thus, while the study of fully coupled system-clock dynamics 
will be reported elsewhere, here corresponding 
backreaction effects are assumed to be small, which characterizes a 
good clock. In a selfconsistent treatment, however, a closed system has   
to include its own clock, reflecting the experience  
of an observer in the universe. 
  
There is no need for the construction of discrete physical time 
in ordinary mechanical systems or 
field theories, where time is an external parameter. 
However, assuming that truly fundamental theories 
will turn out to be diffeomorphism invariant, adding further the requirement that  
observables be quasi-local, then 
such an approach seems natural, which promises to lead to quantum mechanics as an 
emergent description or effective theory on the way.     

\section{From discrete time to evolution of primordial states}
Introducing a functional description of classical mechanics, similarly as in 
Refs.\,\cite{Gozzi1,Gozzi2}, one recognizes the primordial {\it state}: 
\begin{equation}\label{state} 
\langle\tau ,\pi_a|t;t_0\rangle\equiv 
\int\mbox{d}\tau'\int_HD\Phi\exp [
i\int_{\tau'}^{\tau +t}\mbox{d}\tau''L_J
-S(\tau'-t_0)+i\pi_a\varphi^a(\tau +t)] 
, \end{equation} 
and, similarly, the (complex conjugated) {\it adjoint state},   
as useful basic entities for our present purposes \cite{I}; 
$_H$ stands for the presence of the Hamiltonian 
constraint and $\pi_a$ arise from exponentiating $\delta$-functions involving 
$\varphi^a$ at a fixed time; $t_0$ is the initial time. Furthermore,  
$D\Phi\equiv D\varphi D\lambda DcD\bar c$ indicates the functional 
integration over all fields which enter the effective Lagrangian: 
\begin{equation}\label{L} 
L_J\equiv\lambda_a\Big (\partial_\tau\varphi^a-\omega^{ab}\partial_bH\Big ) 
+i\bar c_a\Big (\delta^a_b\partial_\tau -\omega^{ac}\partial_c\partial_bH\Big )c^b
+J_a\varphi^a
\;\;, \end{equation} 
where $J_a$ denotes an external source, $c^a,\bar c^a$ are anticommuting Grassmann variables, 
$\lambda_a$ is an auxiliary variable, and $\varphi^a,a=1\dots 2n$ denote the classical phase 
space variables characterizing the system. 

Generic states of this form, possibly including 
additional weights for different initial conditions of the 
paths contributing in Eq.\,(\ref{state}), are employed to calculate all physical 
(time dependent) {\it observables} of the classical system.  
Considering observables 
which are function(al)s of the phase space variables $\varphi$, we define and calculate: 
\begin{eqnarray} 
\langle O[\varphi ];t\rangle&\equiv&\int\mbox{d}\tau\;P(\tau ;t)
O[-i\frac{\delta}{\delta J(\tau )}]C[-i\frac{\delta}{\delta J(\tau )}]\log Z[J]|_{J=0}
\nonumber \\ [1ex]  
&=&Z^{-1}\int\mbox{d}\tau\mbox{d}\pi\;P(\tau )
\langle t|\tau ,\pi\rangle O[-i\partial_\pi ]C[-i\partial_\pi ]\langle\tau ,\pi |t\rangle 
\nonumber \\ [1ex] \label{expect4} 
&=&\langle\Psi (t)|\widehat O[\varphi ]\widehat C[\varphi ]|\Psi (t)\rangle
\;\;, \end{eqnarray} 
where $Z\equiv Z[0]$ is the generating functional and $C$ the projector 
representing the Hamiltonian constraint, which are not discussed here 
\cite{I}, and where all states refer to $J=0$ as well; here: 
\begin{equation}\label{phihat}
\widehat O[\varphi ]\equiv O[\widehat\varphi ]\;\;,\;\;\; 
\widehat C[\varphi ]\equiv C[\widehat\varphi ]\;\;,\;\;\;
\widehat\varphi\equiv -i\partial_\pi 
\;\;, \end{equation} 
in ``$\tau ,\pi$-representation''. In Eq.\,(\ref{expect4}),  
the final result is for a generic state $|\Psi (t)\rangle$ ($J=0$), 
with the scalar product to be evaluated as in the preceding expression. -- 
Thus, classical observables are represented by corresponding function(al)s of a 
{\it momentum operator}. Its expectation value at physical time $t$ appears as a    
quantum mechanical expectation value, refering to the 
considered state and incorporating a  
$P$-weighted average over its ``history'' in proper time $\tau$.

Now, generic states at physical 
times $t$ and $t+T$ can be related to each other, making use of the 
classical Liouville operator 
propagating phase space variables in $\tau$ \cite{I}. Taking 
the conserved Hamiltonian constraint into account, one obtains a discrete physical 
time {\it Schr\"odinger equation}:  
\begin{eqnarray}\label{discrSchroed}
&\;&\langle\tau',\pi'|\Psi (t+T)\rangle 
\;=
\\ [1ex] \nonumber  
&\;&\int\mbox{d}\tau\;P(\tau) 
\exp [-i(\tau'+T-\tau )\widehat{H}_Q(\pi' ,-i\partial_{\pi'})]
\langle\tau ,\pi'|\Psi (t)\rangle
\;\;, \end{eqnarray}
with the emergent {\it Hamilton operator}:
\begin{equation}\label{Hamiltonian} 
\widehat{H}_Q(\pi ,-i\partial_\pi )\equiv -\pi \cdot\omega\cdot
\frac{\partial}{\partial\varphi}H(\varphi )|_{\varphi =-i\partial_{\pi}}  
\;\;, \end{equation} 
for a given classical Hamiltonian $H$; here $\omega$ is the usual symplectic matrix. 

Equipped with the Hamiltonian $\widehat{H}_Q$, together with the operator $\widehat C$ 
projecting out the constraint subspace, the stationary 
state eigenvalue problem related to Eq.\,(\ref{discrSchroed}) can be 
studied. -- Due to the unusual structure of the Hamiltonian $\widehat{H}_Q$, 
one finds in various examples that the generated spectrum is too rich and, in particular,     
includes notorious negative energy states. Thus, at first sight, the emergent models 
seem not acceptable, since they do not possess a {\it stable groundstate}. However, 
it can be shown that a {\it regularization} (by discretization) of the operators involved 
overcomes this problem. 
  
Thus, while general principles guiding the regularization remain to be investigated, 
for an underlying harmonic oscillator model, one finds the quantum oscillator 
as the corresponding emergent model. Use has to be made of the freedom to choose 
an arbitrary phase of the regularized eigenfunctions, in order to recover the 
correct zeropoint energy (in the continuum limit). Three coupled quantum oscillators 
are found for an underlying relativistic particle model, where the coupling 
is a combined effect of the Minkowski metric and the Hamiltonian constraint \cite{I}.   

Instead of embarking to further describe these models here, some general remarks seem in order.  
It is well known from ordinary quantum (field) theory that the harmonic oscillator is peculiar in 
many respects. Therefore, the reader should not be misled by the present results, namely  
that a quantum harmonic oscillator spectrum is obtained from an underlying classical harmonic 
oscillator model. In particular, it may appear as if this corresponds to ``just another quantization method'', 
in line with canonical commutators, path-integral or stochastic quantization, etc. However, 
this is {\it not} the case. 

It seems an accident of the harmonic system 
that the usual quantized energy spectrum results here. 
This is revealed by the fact, already demonstrated in Refs.\,\cite{tHooft01,tHooft02,ES02,E03}, 
that localization with respect to the coordinate $q$ of the underlying classical model 
has nothing to do with localization with respect to the operator $\hat q$, which is  
introduced a posteriori when interpreting the emergent quantum Hamiltonian corresponding 
to said spectrum. Rather, such localized quantum (oscillator) states are widely spread over 
the q-space of the underlying classical model. 
Interestingly, depending on how the continuum limit is defined in a particular model, 
the $\hat p,\hat q$-commutator may suffer corrections \cite{E03}, 
which interpolate between the classical limit (high energy) and usual quantum 
mechanics (low energy).     
  
Therefore, generally, one cannot expect to find the usual 
quantized counterpart of a classical reparametrization 
invariant model in the present approach, based on discrete physical time. 
Here, the {\it states are collections of classical trajectories}, as defined in 
Eq.\,(\ref{state}), and there is not the usual close correspondence, for example, 
via coherent states. 
On the other hand, in this way a necessary element of quantum mechanical {\it nonlocality} 
enters the description of the underlying reparametrization invariant classical system. 
Further related remarks may be found in Refs.\,\cite{tHooft01,tHooft02,Smolin}.  

To put it differently, the {\it classical limit} of emergent quantum 
theories {\it cannot} be expected to give back the underlying model. 
Instead, the emergent quantum model must be seen as coarse-grained large-scale description of an underlying 
deterministic, possibly dissipative classical dynamics. In itself it is to have a classical limit  
in accordance with the familiar correspondence principle.   

\section*{Acknowledgements} 
It is a pleasure to thank the organizers for the kind invitation 
to the workshop  
``Trends and Perspectives on Extensive and Non-Extensive Statistical Mechanics''.


\begin{thebibliography}{00}




\bibitem{I} H.-T.\,Elze, submitted to Phys.\,Rev. D (2003), quant-ph/0306096.  
\note Additional references can be found here and in the related articles of \cite{II}.     

\bibitem{II} {\it Decoherence and Entropy in Complex Systems}, H.-T.\,Elze (ed.), 
Lecture Notes in Physics, Vol.\,633 (Springer-Verlag, Berlin Heidelberg New York, 2003). 

\bibitem{tHooft01} G.\,'t\,Hooft, {\it Quantum Mechanics and Determinism}, in: Proceedings of the Eigth Int. Conf. 
on ``Particles, Strings and Cosmology'', P.\,Frampton and J.\,Ng (eds.) (Rinton Press, Princeton, 2001), p.275, hep-th/0105105.  

\bibitem{tHooft02} G.\,'t\,Hooft, {\it Determinism Beneath Quantum Mechanics}, quant-ph/0212095.

\bibitem{Wetterich02} C.\,Wetterich, {\it Quantum correlations in classical statistics},  
in: \cite{II}, quant-ph/0212031.

\bibitem{Vitiello01} M.\,Blasone, P.\,Jizba and G.\,Vitiello, 
Phys.\,Lett. {\bf A287}, 205 (2001);   
{\it Dissipation, Emergent Quantization and Quantum Fluctuations},  
in: \cite{II}, quant-ph/0301031. 

\bibitem{Mueller} T.S.\,Biro, S.\,G. Matinyan and B.\,M\"uller, 
{\it Chaotic Quantization: Maybe the Lord plays dice, after all?},  
published in: \cite{II}, hep-th/0301131.   

\bibitem{Smolin} F.\,Markopoulou and L.\,Smolin, {\it Quantum Theory from Quantum Gravity}, 
gr-qc/0311059. 

\bibitem{Adler} S.L.\,Adler, {\it Statistical Dynamics of Global Unitary Invariant Matrix Models as 
Pre-Quantum Mechanics}, hep-th/0206120. 

\bibitem{ES02} H.-T.\,Elze and O.\,Schipper, Phys.\,Rev. {\bf D66}, 044020 (2002). 

\bibitem{E03} H.-T.\,Elze, Phys.\,Lett. {\bf A310}, 110 (2003). 

\bibitem{Gozzi1} E.\,Gozzi, M.\,Reuter and W.D.\,Thacker, Phys.\,Rev. {\bf D40}, 3363 (1989).  
 
\bibitem{Gozzi2} E.\,Gozzi, M.\,Reuter and W.D.\,Thacker, Phys.\,Rev. {\bf D46}, 757 (1992). 

\end{thebibliography}
\end{document}